\documentclass[%
 reprint,
 amsmath,amssymb,
 aps,
10pt]{revtex4-2}

\usepackage{graphicx}
\usepackage{dcolumn}
\usepackage{bm}
\usepackage{orcidlink}
\usepackage[normalem]{ulem} 
\DeclareMathOperator{\Tr}{Tr}
\usepackage{float}

\begin{document}

\preprint{APS/123-QED}

\title{Multiple Quantum Many-Body Clustering Probed by Dynamical Decoupling }

\author{Gerónimo Sequeiros$^{1,2}$\,\orcidlink{0009-0007-2259-1322}}
\author{Claudia M. Sánchez$^{1}$\,\orcidlink{0000-0001-5965-4963}}
\author{Lisandro Buljubasich$^{1,2}$\,\orcidlink{0000-0002-4629-0870}}
\author{Ana K. Chattah$^{1,2}$\,\orcidlink{0000-0003-0187-9654}}
\author{Horacio M. Pastawski$^{1,2}$\,\orcidlink{0000-0002-9573-9865}}
\author{Rodolfo H. Acosta$^{1,2}$\,\orcidlink{0000-0003-3945-8273}}%
    \email{r.acosta@unc.edu.ar}
\affiliation{$^1$ Facultad de Matemática, Astronomía, Física y Computación, Universidad Nacional de Córdoba, Córdoba, Argentina}
\affiliation{$^2$ CONICET, Instituto de Física Enrique Gaviola (IFEG), Córdoba, Argentina}


\begin{abstract}
\noindent
The manipulation of quantum information in large systems requires precise control of quantum systems that are out-of-equilibrium. As the size of the system increases, its fragility in response to external perturbations and intrinsic decoherence processes also increases. The degradation of the system response makes accurate measurements a challenging and time-consuming task. However, quantum information lifetime enhancement can be achieved by dynamical decoupling techniques (DD), where an external drive with a frequency much higher than the system's internal evolution renders signal acquisition with decay times greater than 1000-fold. In this study, we demonstrate that the system response during a prethermal period, subject to Floquet control, can be utilized to probe the multiple quantum evolution of dense and highly connected spin systems. This approach exhibits an enhanced sensitivity at a reduced experimental time. The enhanced signal-to-noise ratio achieved enabled the use of numerical inversion strategies to model the evolution of the excited multiple quantum coherences, which describe the number of correlated spins within a cluster. We observed for the first time, to the best of our knowledge, that the increase in the number of correlated spins with multiple quantum evolution is accompanied by an increase in the distribution of spin cluster sizes, which follows a quadratic law.
\end{abstract}

\maketitle

\begin{figure}[hbt]       
    \includegraphics[width=\columnwidth]{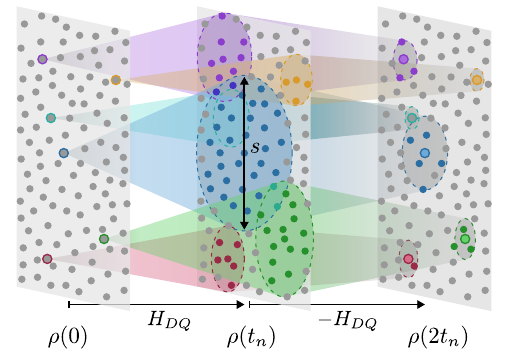}
    \caption{\textbf{Scheme of initial excitations,  scrambled states, and time reversal.} The left panel exemplifies various spin local excitations through colored dots. The central panel shows the scrambled excitations into spin clusters of different sizes,  \(s\).  The time reversal dynamics is only partially successful (third panel). 
    This procedure provides  OTOCs that characterize the distribution of clusters in the central panel.}  
    \label{fig:diagrama}
\end{figure}

The thermalization of any local excitation in a large many-spin system is preceded by reversible spreading with Hamiltonian dynamics, which occurs on a time scale of $ T_2$. Thus, the information concerning the excitation becomes encoded in many-body correlations within spin clusters of increasing size. The description of these correlations requires an exponentially growing number of states, among which the information becomes scrambled, as illustrated in Fig. \ref{fig:diagrama}. The states sketched in the central panel are in principle not observable; however, the information can be encoded by tailored measurement protocols.
The propagation velocity of this wavefront can be characterized by the the temporal increase in the square commutator of two local operators which initially commute.. This defines an out-of-time order commutator (OTOC) \cite{RoSw12,Sw18} which is connected to various fundamental problems, ranging from superconductivity in dirty metals\cite{LaOv69} and spin networks\cite{MuPi86,Khithin97}, to quantum chaos\cite{Mald20} and a route to address the information paradox in black holes\cite{Ki17}. A considerable effort has been  invested in determining the growth of OTOC that characterizes scrambling in specific systems. This growth depends on the specific characteristics of the Hamiltonian, the lattice, and the spatial distance of the observables that are both excited and detected (i.e., the operators in the OTOC). 
Experiments involve a time-reversal procedure and local measurements, schematized in the third panel of Fig. \ref{fig:diagrama}, by the shaded zones around the initial excitations..


Further control of the quantum dynamics is achieved by the imposition of a periodic external drive that interrupts and modifies the information's natural spreading. 
The requirement for stopping the system dynamics is that the driving period should be much shorter than the system's interaction timescale \cite{Goldman2014_PRX,Weidinger2017_SciRep}. This leads to a Floquet dynamics, where the effective Hamiltonian time scale $T_2$  can be increased up to infinity. In this case, the decay of the excitation depends on $\Sigma$, the small interactions and environmental effects omitted from the engineered effective Hamiltonian. 
This defines an irreversibility time scale $T_3$, in terms of the natural perturbation time scale $T_\Sigma$. On the other hand, maximizing the Hamiltonian dynamics is equivalent to a relative slowing down of the rate $1/T_\Sigma$. Many different experiments concurrently reach a perturbation-independent irreversibility time scale of about $T_3\approx 6T_2<T_\Sigma$ \cite{Sanchez2020,Sanchez2022,Sanchez2023}.  This fact makes evident it clear that once the information is encoded in to this scrambled, pre-thermalized state, it becomes so extremely fragile that even an infinitesimal uncontrolled process renders it irreversible. In this regard, the study of scrambling and pre-thermalization dynamics is of crucial importance for the foundations of physics\cite{BrSuss18}, as well as for technological development. Realistic numerical simulations are immediately out of reach. All this requires new computational strategies\cite{XuSw19,LoZgPa21,McCar24}, statistical estimations\cite{DoAl21,ScYao23,ZhSw23}, and quantum simulations\cite{YaSi20,Gr_Li22,McCar24}.

In the field of nuclear magnetic resonance (NMR), particularly solid-state NMR, the periodic drive mentioned above takes the form of radio-frequency pulses. The rotation angles, inter-pulse spacing, and periodicity of these pulses precisely determine the quantum dynamics of the systems under study. 
Dynamical decoupling (DD), a particular category of Floquet driving, has been widely used to preserve states as coherent superpositions\cite{AlvSut16,Ashok_PRL2021}. 
Originally conceived for weakly interacting quantum systems coupled to an external bath, the DD techniques were designed to modulate the system-bath interaction and prevent decoherence. More recently, these sequences have been extended to systems where internal interactions dominate over bath coupling, in particular in closed dilute systems such as $^{13}$C nuclei in diamond where an extension of state lifetimes proportional to the strength of dipolar coupling has been observed \cite{Ashok2020,Ashok2023_NatPhys_DTC}.
In this article, we use DD as a detection block aimed to increase the signal-to-noise ratio, thereby enabling faster and more precise experiments. Specifically, we use an OTOC experiment to characterize a highly interconnected nuclear spin system evolving under a double-quantum Hamiltonian. 
This allows us to obtain a detailed account of how the spreading occurs  and to give a definitive assessment of the butterfly velocity, greatly improving previous experimental estimations. Our results confirm that the size of the largest cluster grows linearly with time \cite{Sanchez2023,Sanchez2020,Li2024_NatPhys}.
\begin{figure*}[hbt]
    \includegraphics[width=2\columnwidth]{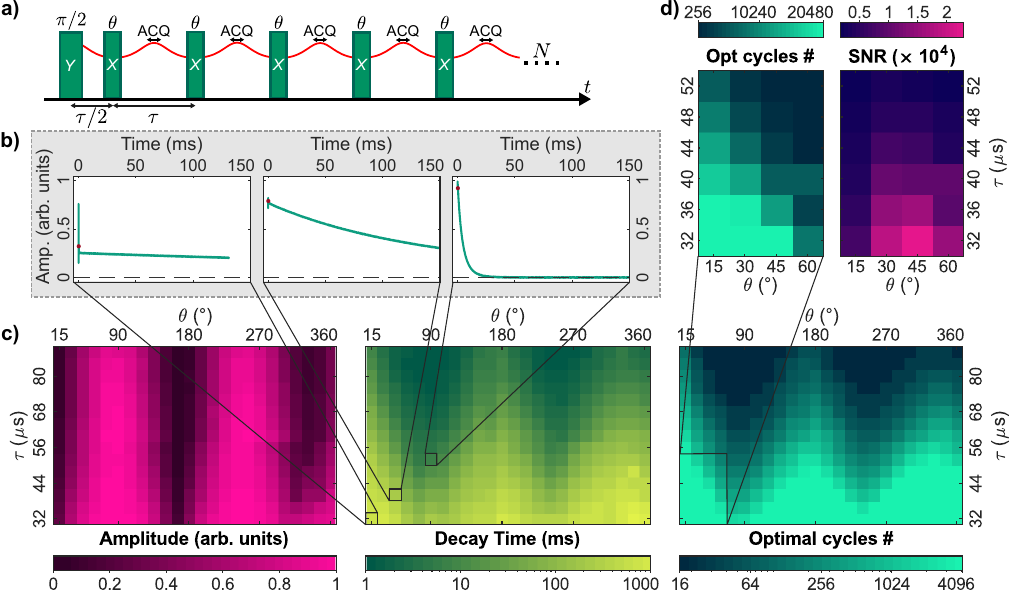}
    \caption{\textbf{Dynamical Decoupling sequence optimization:} \textbf{a)} Pulse sequence diagram. A $\pi/2$ rotation on the $Y$ axis of the rotating frame is followed by a train of $N$ rotations ($\theta$) with frequency $1/\tau$. The system evolution is probed at the center of the excitation cycles. \textbf{b)} Signal evolution for different parameter settings showing variations of initial transitory oscillation, amplitude, and decay times. Red dots indicate the first data point used for fittings after the initial oscillating transient. \textbf{c)} Combined exponential amplitudes, characteristic decay times, and optimal number of cycles as a function of $\tau$ and $\theta$.  \textbf{d)} Reduced parameter map showing optimal number of Floquet cycles and respective SNR with an increased $N=20480$ cycles.}
    \label{fig:1}
\end{figure*}

A scheme of the DD pulse sequence used in this work is depicted in Fig. \ref{fig:1}a.
An initial $\pi/2$ pulse aligned with the $Y$-axis of the rotating frame, which drives the system out of equilibrium, is followed by a train of $N$ excitation pulses with period $\tau$ and generic rotation angle $\theta$, all aligned with the $X$-axis of the rotating frame.
The evolution of the spin system during the sequence is monitored by means of acquisition windows placed between successive $\theta$ pulses. 
The experiments presented in this article were carried out in plycristalline adamantane. Due to the rapid molecular tumbling, each adamantane molecule can be considered as a single spin-$1/2$, located in a face-centered cubic lattice, affected only by intermolecular dipole-dipole interactions. Therefore, the hydrogen nuclei in adamantane ara a good example of a highly interconnected spin system.
The secular part of the dipolar Hamiltonian with respect to the dominant Zeeman interaction is
$\mathcal{H}_{\text{d}}^{zz}=\sum_{i<j}d_{ij}(3I_i ^zI_j^z-
\textbf{I}_i \cdot \textbf{I}_j) $, where $I_i^{\alpha}$ ($\alpha=x,y,z$ represents the $i-$spin operator, and the dipolar coupling strengths $d_{ij}$ decrease with the internuclear distance.

Due to the complexity of the interactions in many-body systems, we resort to an empirical examination of the dependence of the system's evolution on a different set of parameters.
By changing the Floquet frequency and the rotation angle of the periodic excitation, two effects are readily observed (see Fig. \ref{fig:1}b); both the initial signal intensity and the decay rates depend on $\tau$ and $\theta$. 
The signal exhibits an initial oscillatory transient characteristic of periodic excitations \cite{Acosta2003}, lasting up to $8$ Floquet cycles depending on the parameter choice (peak oscillation shown in the left panel of Fig. \ref{fig:1}b). 
The signal response for a sweep of the parameters $\tau$ and $\theta$ is shown in Fig. \ref{fig:1}c. Its evolution shows a bi-exponential decay, consisting of both a fast and a slow decaying exponential, resembling prethermatization regimes and thermalization processes observed in weakly coupled systems \cite{Santos2021,Cappellaro2021_NatPhys,Mori2023_AnnRev,Mori2023_AnnPhys}.
The left panel of \ref{fig:1}c shows the fitted total amplitudes (i.e. the extrapolation of the total signal to $t=0$), while the central panel shows the fitted characteristic decay time for the slowly decaying exponential, that describes the total signal behavior.
Longer decay times correspond to shorter $\tau$ values. For $\theta$, an oscillatory behavior is observed where higher signal intensities correspond to shorter decay times.

The optimal number of Floquet cycles to maximize the SNR was determined empirically, as shown in the right panel of Fig. \ref{fig:1}c. 
The saturated scale in the last plot indicates that more than four thousand transients are needed to fully exploit the long-lasting evolution; however, heating the excitation coil in successive Floquet cycles imposes an experimental limitation.
In Fig. \ref{fig:1}d we show a reduced parameter map where the SNR is determined for up to $2048$ Floquet cycles. 
We find that for this given number of cycles the choice of $\theta=45^{\circ}$ is optimal, while the trend for $\tau$ suggests that shorter values are more favorable for the spin system considered in this work.

 \begin{figure*}[hbt]         
    \includegraphics[width=2\columnwidth]{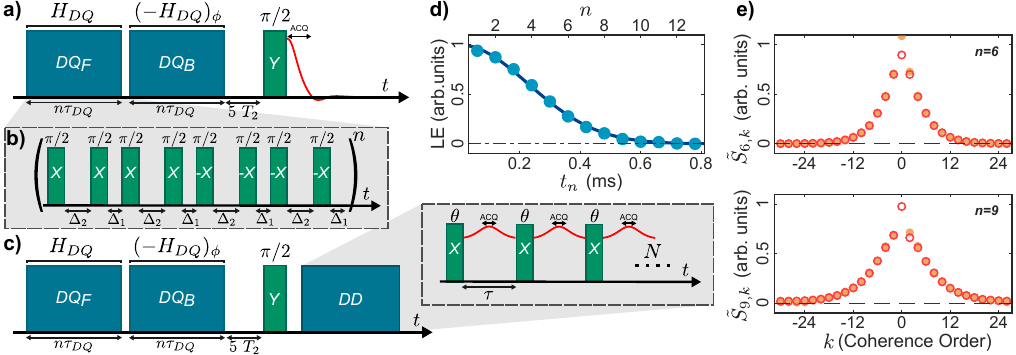}
    \caption{\textbf{Multiple Quantum Coherences:} \textbf{a)} Diagram of a MQC experiment. The first DQ$_F$ sequence is composed of $n$ consecutive DQ blocks, acting over the system as the effective Hamiltonian $H_{DQ}$, then the DQ$_B$ sequence reverts this evolution with an effective Hamiltonian $(-H_{DQ})_\phi$. After transverse relaxation, a readout $\pi/2$ pulse is used to acquire an FID. \textbf{b)} Diagram of the DQ pulse sequence utilized. Each block consists of $8$ pulses separated by the free evolution periods indicated in the figure ($\Delta_1=3\ \mu\text{s}$ and $\Delta_2=8\ \mu\text{s}$), accounting for a total evolution period of $\tau_{DQ}=60\ \mu\text{s}$. \textbf{c)} Implementation of the DD sequence to probe the MQC experiment. \textbf{d)} Loschmidt Echo amplitude for increasing DQ evolution obtained from DD sequence with fixed parameters. \textbf{e)} Distributions $\tilde{S}_{n,k}$ of excited even coherence orders obtained by FID (solid circles) and DD (hollow circles) acquisitions corresponding to $n=6$ and $n=9$ blocks of DQ excitation.}
    \label{fig:2}
\end{figure*}


The performance of the DD scheme as a detection block is evaluated in a series of experiments that involve the correlation of a large number of nuclear spins. NMR has long been a platform for controlling spin dynamics by engineering pulse sequences that modulate the system interactions.
In particular, the Double Quantum (DQ) sequence was designed to obtain an effective Hamiltonian $H_{DQ}$, 
    \begin{equation}
        H_{DQ} = -\tfrac{1}{2}\sum_{i<j} d_{ij} (I_i^+I_j^+ + I_i^-I_j^-)
    \end{equation}
\noindent which is the zeroth order of the Magnus expansion in the Average Hamiltonian Theory \cite{Hb76}.  
Evolution under this Hamiltonian enables pairs of spins with the same orientation to flip, leading the system to develop coherent states involving an increasing number of spins \cite{BMGPi85, BPi86}. These coherent states, characterized by even orders of coherence, form clusters of spins that grow over time. The size of these clusters can be probed in a Multiple Quantum Coherence (MQC) experiment, which involves the measurement of different time-reversed projections of DQ evolved states.
A diagram of the MQC experiment is depicted in Fig. \ref{fig:2}a.
In this experiment, the system starts from a thermal equilibrium state $\rho(0) \propto I^z$ (here $I^z=\sum_i I_i^z$), and evolves under the effective Hamiltonian $H_{DQ}$ by applying $n$ blocks of DQ$_F$ (see Fig. \ref{fig:2}b), for a total evolution time of $t_n=n\tau_{DQ}$.
Next, $n$ blocks of DQ$_B$ are applied to reverse the initial evolution with a phase difference, evolving the system under the effective Hamiltonian $(-H_{DQ})_{\phi}= e^{-i\phi I^z} (-H_{DQ}) e^{i\phi I^z}$ for the same duration $t_n$. The propagator accounting for the total evolution-reversion period is $U_\phi(2t_n) = e^{-i\phi I_z}e^{it_nH_{DQ}}e^{i\phi I_z}e^{-it_nH_{DQ}}$ and the resulting system's density matrix is given by $\rho_\phi(2t_n) = U_\phi(2t_n)\rho(0)U_\phi^\dagger(2t_n)$.
Finally, a Free Induction Decay (FID) signal is acquired with a readout pulse after a filter period of free evolution, included to allow any spurious transverse magnetization to average out. The signal measured for a given reversion phase $\phi$ and number of DQ blocks $n$, $S_{n,\phi} \propto \Tr\left\{I_z \rho_\phi(2t_n)\right\}$, represents a generalized echo \cite{Lozano2024, GJaPaWi12}. 

Many-body systems are sensitive to decoherence, and experimental imperfections hinder perfect evolution reversal. These imperfections can be quantified by the Loschmidt Echo (LE) $S_{n,0}$ which is the resulting signal after forward and backward evolution with no phase difference $\phi=0$, and represents the ability to recover the initial state of the system \cite{Sanchez2020}.
As the evolution time $t_n$ increases, the size of the cluster of correlated spins grows and the system becomes more sensitive to perturbations and imperfections during the DQ evolution. This leads to a degradation of the final state, making it harder to extract information from these experiments. The DD sequence as a detection block is aimed to improve the SNR in the acquisition of highly deteriorated states, and is implemented as shown in Fig. \ref{fig:2}c. The signal degradation, measured through the LE, is shown in Fig. \ref{fig:2}d, where the decrease in signal for large correlated spin clusters is evident. Improving the  acquisition for long evolution times $t_n$ enables better probing of the evolution of coherence clusters  in the system.
The signal $S_{n,\phi}$ can be expressed in terms of the coherence orders of the density matrix that develop during the first period of DQ evolution:
\begin{equation}
    S_{n,\phi} = \sum\limits_k e^{-i\phi k}\sum\limits_r|\rho_{r,r+k}(t_n)|^2 = \sum\limits_k e^{-i\phi k} \tilde{S}_{n,k}
\end{equation}
Here $\tilde{S}_{n,k}$ represents the distribution of coherences of order $k$ (even values only) developed during the forward evolution (DQ$_F$), and is the inverse Fourier Transform  of the measured signals $S_{n,\phi}$. 
The MQC experiment requires a full rotation on the $\phi$ dimension, with the step size determining the maximum even order of coherences to be probed. Since these dimensions are related by a Fourier Transform, exploring higher orders of coherence requires incrementing $\phi$  at smaller angular steps, thus involving more experimental time.
The distributions of coherences $\tilde{S}_{n,k}$ of Fig. \ref{fig:2}e show data acquired  using both FID and DD acquisition schemes for two different block numbers, $n=6$ and $n=9$. The results show that the acquisition with DD reproduces the behavior from the original experiment for these number of DQ blocks. These distributions of coherences contain information about $s$, the sizes of the spin clusters that have developed at the evolution times. 
The number of correlated spins corresponds to the second moment of the distribution of coherences, and can be expressed as an OTOC \cite{Lozano2024, GJaPaWi12, Sw18}: 
$ s \propto   \Tr\left\{\left[\hat{I}^{z},\hat{I}^{z}(t_n)\right]\left[\hat{I}^{z},\hat{I}^{z}(t_n)\right]\right\}$, where $I^z(t_n)$ is used to represent the operator $I^z$ following the evolution of the system for a period of time $t_n$.

\begin{figure*}[hbt]        
    \includegraphics[width=2\columnwidth]{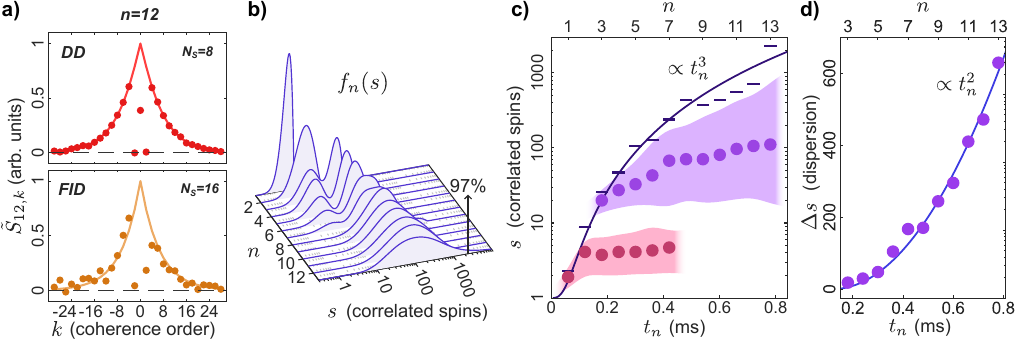}
    \caption{\textbf{a)} Comparison of the distribution of coherences obtained by FID and DD for $n=12$ DQ blocks. Solid lines are guides to the eye. Here the improvement through DD acquisition becomes evident as it gives much more defined distributions. \textbf{b)} Distribution of cluster sizes $f(s)$ for different number of cycles in the DQ evolution. \textbf{c)} Evolution of the distinct populations of $f(s)$. Maximums of the distributions are depicted in circles and their respective FWHM encloses the background areas for each population. The $97\%$ cumulative values (arrow in $f(s)$) for each distribution are represented by horizontal bars, which align with the solid curve representing cubic growth. \textbf{d)} Dispersion ($\Delta s$) of the correlated spin cluster size corresponding to the FWHM of the larger population on c). We observe that $\Delta s$ grows quadratically $\propto t_n^2$ as represented by the solid curve. }
    \label{fig:3}
\end{figure*}

One conventional way to compensating for low SNR is to average multiple experimental realizations (scans). This is where the DD sequence acquisition shows its potential, as it allows for more defined results and significantly reduces the number of scans $N_S$ compared to the case of FID acquisition.
Fig \ref{fig:3}a shows a comparison of the coherence distribution obtained through both FID and DD acquisition for $n=12$ blocks of DQ$_F$ evolution, $t_{12} = 720\ \mu$s.
For this very weak signal, the optimization shown in Fig. \ref{fig:1} fails because the signal evolution rapidly reaches the noise level, as observed in the LE experiment of Fig. \ref{fig:2}d . In this particular case, the SNR was maximized by reducing the number of Floquet cycles to 256 as well as the pulse separation to $\tau = 10\ \mu$s.
Data acquired with the DD sequence is sharper and gives a well-defined distribution with $N_S=8$, and is still sharper than the distribution obtained with $N_S=16$ using an FID acquisition (see Fig. \ref{fig:3}a). We estimate that $\sim 22$ times more scans are needed for the FID to achieve the same SNR obtained with the DD sequence. 
The most accepted model for determining the distribution of coherences was proposed by Baum et al. \cite{BMGPi85,BPi86}, where a Gaussian function is used to describe the data, and the number of correlated spins is determined from the width of this Gaussian. However the Gaussian function often fails to represent the data correctly for long evolution times. Different strategies have been implemented \cite{Khithin97}, such as introducing a stretched exponential \cite{Lacelle93} or describing the spin clustering as two separate clusters of different sizes \cite{Sanchez2016,Sa+Ch14}. Here we describe $\tilde{S}_{n,k}$ for each value of $n$, as a weighted distribution of Gaussian functions with varying widths:

\begin{equation}
    \tilde{S}_{n,k} = \sum_j \exp\left( -\frac{k^2}{s_j}\right) f_n(s_j)+ \epsilon_{n,k},
\end{equation}
\noindent where $\epsilon_{n,k}$ is a vector representing the noise for each MQC distribution, the exponential term corresponds to the Gaussian kernel, and $f_n(s)$ is the probability of finding a cluster of size $s$ at a given time $t_{DQ}$.
Finding $f_n(s)$ given $\tilde{S}_{n,k}$ is an ill-posed problem that can be solved numerically. Provencher \cite{Provencher82} implemented parsimony using a regularization method known Tikhonov regularization \cite{Tycho79}, which stabilizes the fitting by smoothing the solutions. In this work, we use a one-dimensional algorithm introduced by Teal and Eccles \cite{Teal_2015}. Figure \ref{fig:3}b shows the cluster size distributions obtained as a function of the number of DQ blocks (cycles) applied. Initially, a single distribution with a population (maximum) of few correlated spins is observed, which decays very rapidly over time. Bi-modal distributions, with a second population of large clusters, emerge for $n = 3$, whose maximum and width increase with time. The number of correlated spins increases systematically until, for a large number of cycles, the experimental procedure renders the Loschmidt Echo signal, Fig.\ref{fig:2}d, undetectable. The maxima of the bi-modal distributions are represented by the solid circles in Fig. \ref{fig:3}c. For small $n$ values (short times) the double quantum dynamics produces a collection of small clusters of similar size. Clusters of increasing size are observed to become more dispersed over time, where the width of the distributions is represented by shadows. Additionally, the propagation of the fastest cluster can be determined from the distributions of Fig. \ref{fig:3}b. As a reference value we use $97\%$ of the individual cumulative values (arrow in $f(s)$) for each distribution. The propagation front is represented as horizontal lines in Fig. \ref{fig:3}c, which are observed to grow as $t^3$, in accordance with the ballistic dynamics previously observed for the double quantum Hamiltonian in a uniform 3D network such as adamantane\cite{Sanchez2023}.

Finally we focus our attention on the width of the distributions, considering for short evolution times only the one that represents the group of larger clusters. We must keep in mind that the evolution under the forward and backward multiple quantum Hamiltonians not only creates spin clusters of correlated spins that grow in size, but it also keeps on connecting spins corresponding to small cluster sizes. The distribution of cluster sizes, which is represented in the central panel of Fig. \ref{fig:diagrama}, could not be determined so far with the usual  fitting methods. The use of the numerical inversion used in this work enabled us to observe this phenomena for the first time, which has a quadratic dependence on the evolution time as shown in Fig. \ref{fig:3}d. Even though we do not yet have a model to describe the underlying spin dynamics that account for this observation, we believe that the empirical observation itself can be useful to the large community devoted to the study of many body systems.

In summary, we have designed a detection toolbox for the acquisition of signals describing the dynamics of many-body systems out-of-equilibrium. The dynamical decoupling evolution under successive Floquet cycles for strongly interacting spin systems renders the addition of thousands of transients that boost the sensitivity, enabling rapid detection of large spin clusters involving hundreds of coupled spins whose signal is degraded by the Loschmidt echo. The improved signal-to-noise ratio achieved is sufficient to apply numerical inversion methods, which are extremely noise sensitive, to model de distribution of connected spins forming clusters of different sizes through the evolution of MQ coherences. The observation of new dynamics is a subject of ongoing studies for which we yet lack a model to describe.

\bibliography{DD_DQ_Gero}
\bibliographystyle{unsrt}

\end{document}